\documentclass{emulateapj}
\def\be{\begin{equation}}
\def\ee{\end{equation}}

\def\bf{\begin{figure}}
\def\ef{\end{figure}}

\def\bea{\begin{eqnarray}}
\def\eea{\end{eqnarray}}

\newcommand{\prob}[5]{\ensuremath{f_{#1}(S_{#2}|\omega_{#3},S_{#4},\omega_{#5})}} 

\newcommand{\probm}[5]{\ensuremath{f_{#1}(S_{#2},\omega_{#3};S_{#4},\omega_{#5})}} 

\newcommand{\probmerge}[3]{\ensuremath{f(S_{#1}\rightarrow S_{#2};\omega_{#3})}} 

\newcommand{\boldm}[1]{\ensuremath{\mbox{\boldmath $#1$}}}

\usepackage{graphicx}
\usepackage{amsmath}
\begin{document}
\title{Dark matter halo merger and accretion probabilities in the excursion set formalism}
%
%
\author{Esfandiar Alizadeh}
\affiliation{\itshape Department of Physics, University of Illinois at Urbana-Champaign, 1110 West Green Street, Urbana, IL 61801}
\author{Benjamin Wandelt}
\affiliation{\itshape Department of Physics, University of Illinois at Urbana-Champaign, 1110 West Green Street, Urbana, IL 61801}
\affiliation{\itshape Department of Astronomy, University of Illinois at Urbana-Champaign, 1002 West Green Street, Urbana, IL 61801}
\affiliation{\itshape Center for Advanced Studies, University of Illinois at Urbana-Champaign, 912 West Illinois Street, Urbana, IL 61801}
\submitted{}
%
%
%
\begin{abstract}

The merger and accretion probabilities of dark matter halos have so far only been calculated for an infinitesimal time interval. This means that a Monte-Carlo simulation with very small time steps is necessary to find the merger history of a parent halo. In this paper we use the random walk formalism to find the merger and accretion probabilities of halos for a finite time interval. Specifically, we find the number density of halos at an early redshift that will become part of a halo with a specified final mass at a later redshift, given that they underwent $n$ major mergers, $n=0,1,2,...$ . We reduce the problem into an integral equation which we then solve numerically. To ensure the consistency of our formalism we compare the results with Monte-Carlo simulations and find very good agreement. Though we have done our calculation assuming a flat barrier, the more general case can easily be handled using our method. This derivation of finite time merger and accretion probabilities can be used to make more efficient merger trees or implemented directly into analytical models of structure formation and evolution.

\end{abstract}
\subjectheadings{cosmology:theory-dark matter}
\maketitle
\pagestyle{empty}
%
%
\section{Introduction}  \label{sec:intro}

\indent
Since the introduction of the spherical collapse model of dark matter halos by Press and Schechter \citep{Press} and its generalizations such as extended Press-Schechter (EPS) \citep{Bond} and ellipsoidal collapse \citep{Sheth} it has been used extensively in the cosmological literature as a fast and accurate method to quantify the distribution of collapsed dark matter objects in the universe. This in turn is the backbone of semi-analytic theories of galaxy formation and evolution \citep{Cole}. Modified versions of the excursion set formalism underlying EPS have also found application in different contexts such as ionized bubble growth in the early universe \citep{Furlanetto}. The advantages of this method compared to direct N-body simulation include its superior speed, which allows the exploration of large ranges of 
parameter space, and redshift range than is currently accessible to N-body simulations. 

In the EPS formalism one finds the probability $P(M_1,z_1|M_2,z_2)dM_1$ which gives the probability of a point mass being part of a halo with mass in between $M_1$ and $M_1+dM_1$ at redshift $z_1$ given that it was (or will be) part of a halo with mass $M_2$ at redshift $z_2$. During this redshift interval it could merge with any number of halos whose masses add up to $|M_1 - M_2|$. However, from the viewpoint of galaxy formation, accretion of small halos into the larger one do not have the ability to change the evolution of the galaxy or galaxies inside it. Indeed, it is assumed that only at \emph{major mergers} in which the mass of the merged halo satisfies a condition to be large enough, have this ability. It is therefore interesting to ask the question: given the criterion for a major merger, is it possible to find an analytical result to describe the progenitor distribution of a parent halo, based on how many times they have undergone a major merger? For example, what is the number density of halos at redshift, say, $z=1$ which underwent $n$ major mergers, $n=0,1,2,...$, and eventually ended up being a galaxy halo at the present time. 

Here we present an analytical method based exclusively on excursion set assumptions to find these number densities. We reduce the problem to an integral equation which can then be solved numerically. Our derivation is given in section \ref{sec:theor}. We show in section \ref{sec:compare} that Monte-Carlo (MC) simulations agree with the results of our semi-analytic model.

It is also possible to cast the integral equation into a scale invariant form. This can be a great advantage since we just need to solve the integral equation once and scale the result to find the general formula. This is done in section \ref{sec:scaled}. We discuss how the method presented in this paper can be generalized and conclude in section \ref{sec:discussion}.

 A few appendixes are added for further clarification. Appendix \ref{sec:halo} gives a brief introduction to the excursion set formalism and defines our notation. In Appendix \ref{sec:num} we give a straightforward method to solve the integral equation.
 Finally in Appendix \ref{sec:monte} we describe very briefly how we implemented our Monte-Carlo simulation. 
 
\section{The Accretion probability}  \label{sec:theor}

 The main task of this paper is to find an analytical result for $\prob{acc}{2}{2}{1}{1}$, the probability that a random walk starting from $(S_1,\omega_1)$ has its first upcrossing between $S_2$ and $S_2+dS_2$ at the barrier height $\omega_2$ given that it never had a jump larger than $M_{res}$ between $\omega_1$ and $\omega_2$. Here $S=\sigma^2(M)$ is the rms mass fluctuation inside spheres of mass $M$, and $\omega$ is a monotonically increasing function of redshift which for the case of an EdS universe takes the familiar form of $1.68 \times(1+z)$ \footnotemark[$\dagger$] \footnotetext[$\dagger$]{For further explanation of excursion set formalism, see Appendix \ref{sec:halo}}. In order to accomplish this we let a number of random walks start from $(S_f,\omega_f)$ (see fig. \ref{fig:randomwalk}), where in this paper $S_f$ and $\omega_f$ denote $\sigma^2(M_f)$ and $\omega(z_f)$ respectively, and so forth for other subscripts. Setting a barrier at $\omega_i>\omega_f$, that is at earlier times, we know the fraction of random walks that have their first upcrossing in the interval $(S_i,S_i+dS_i)$, regardless of whether they accrete or merge, is (see equation \ref{equ:ptotapp}):
 
 \begin{figure}
\begin{center}
\includegraphics[width=84mm]{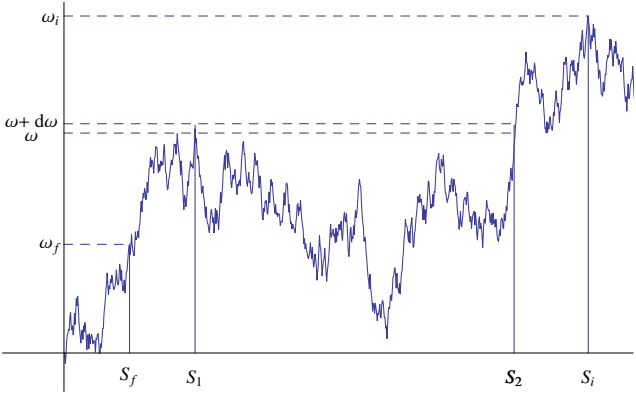}
\end{center}
\caption{This figure shows a sample trajectory for a point that belongs to a halo of mass $S_f$ at time $\omega_f$ in a $S$ vs. $\omega$ diagram. From this figure we can find places where the halo to which this point belongs splits into two smaller halos of masses larger than $M_{res}$ by finding the places where the first upcrossing of the random walk has a jump larger than $\Delta S=S(M)-S(M-M_{res})$ when the barrier height is increased from $\omega$ to $\omega + d \omega$. In this figure the mass resolution is such that there is just one jump large enough to be considered a merger, and occurs where the first upcrossing jumps from $S_1$ to $S_2$. The rest of the time the halo will just accrete masses smaller than $M_{res}$. We want to find the fraction of random walks that start from $(S_f,\omega_f)$ and will have their first upcrossing between $S_i$ and $S_i+d S_i$ at the barrier height $\omega_i$ and will have one and only one large enough jump to be considered as merger. We first find the fraction of the random walks that, starting from $(S_f, \omega_f)$, just accrete mass and end up somewhere in the mass range $(S_1,S_1 +d S_1)$ at the barrier height $\omega$. Then we take the random walks that passed the previous test and find the fraction of them that have a merger from $S_1$ to some $S_2>S(M(S_1)-M_{res})$ during an infinitesimal time interval $d \omega$. Finally, from these random walks we take the fraction that, starting from $(S_2,\omega)$, only accrete and end up in the mass range $(S_i,S_i+dS_i)$ at the barrier height $\omega_i$. The outcome of this is the equation \ref{equ:1M}. To find the total fraction of the random walks that have one and only one merger we need to integrate over intermediate values of $S_1$, $S_2$ and $\omega$, keeping in mind that $S_2$ cannot come closer to $S_1$ than what the mass resolution allows us. This give us the equation \ref{equ:1merge}. }
\label{fig:randomwalk}
\end{figure}

\bea
 &   &\prob{tot}{i}{i}{f}{f}\,dS_i \nonumber \\
 & = & \frac{(\omega_i -\omega_f)}{(2\pi)^{1\!/2}(S_i-S_f)^{3/2}}\exp\left[-\frac{(\omega_i-\omega_f)^2}{2(S_i-S_f)}\right] \,dS_i \label{equ:ftot}
\eea
 The next step is to notice that the fraction of random walks that start from $(S_f,\omega_f)$ and have their first upcrossing between $S_i$ and $S_i+d\,S_i$ at the height $\omega_i$ and have one and only one jump from $S_1$ to $S_2$ during the interval $\omega$ and $\omega+d\omega$ is:
 \bea
 &&\prob{acc}{1}{}{f}{f} \,dS_1 \times\ \probmerge{1}{2}{} \,dS_2 \,d\omega \nonumber \\
   &\times& \prob{acc}{i}{i}{2}{} \,dS_i , \label{equ:1M}
 \eea 
where $\probmerge{1}{2}{}\,dS_2 \,d\omega$ is the probability that a random walk will have a sudden jump from $S_1$ to somewhere between $S_2$ and $S_2+dS_2$ in an infinitesimal interval $d\omega$. Its form for the Spherical collapse model is given by equation \ref{equ:pmerge}.
 
 Now, if we integrate over all $S_1$, $S_2$ and $\omega$ keeping in mind that $M(S_1)-M(S_2)$ must be larger than $M_{res}$ to be classified as a merger, we find the fraction of walks that have undergone one and only one merger to be:
 
 \bea
&  &\prob{1merger}{i}{i}{f}{f}=\int_{\omega_f}^{\omega_i} \!\,d\omega \int_{S_f}^{S_i-\Delta' (S_i)} \! \,dS_1  \int_{S_1+\Delta (S_1)}^{S_i} \,dS_2 \nonumber \\
&  & \prob{acc}{1}{}{f}{f} \probmerge{1}{2}{} \prob{acc}{i}{i}{2}{} \nonumber \\
&   &    \label{equ:1merge}
 \eea
 
 Here $\Delta (S_1)=\sigma^2(M(S_1)-M_{res})-S_1$ is the closest distance that $S_2$ can be brought to $S_1$ and still have a jump large enough to be considered as a merger. Analogously define $\Delta' (S_i)=S_i-\sigma^2(M(S_i)+M_{res})$. 
 
 We can similarly find the fraction of random walks that have undergone two and only two mergers, $\prob{2mergers}{i}{i}{f}{f}$. We get an expression like equation \ref{equ:1merge} but with six integrations, and higher terms involve more integrations yet.
 Adding all of these terms must give:
 \bea
 &  &\prob{tot}{i}{i}{f}{f}=\prob{acc}{i}{i}{f}{f}+ \nonumber \\
 &  &\prob{1merger}{i}{i}{f}{f}+\prob{2mergers}{i}{i}{f}{f}+\cdots \nonumber \\
 &  &\label{equ:summ}
 \eea
 which says that the fraction of random walks starting from $(S_f,\omega_f)$ and passing through a barrier at $\omega_i$ for the first time between $(S_i,S_i+dS_i)$, no matter what happened during their journey, is equal to the sum of the fractions of walks which underwent no mergers, one merger, two mergers, etc. 
 
Putting the integral formulae for $\prob{1merge}{i}{i}{f}{f}$ and higher order terms into equation~\ref{equ:summ}, we will find an integral equation for the unknown $\prob{acc}{i}{i}{f}{f}$. However, this equation written in this form is not computationally tractable since higher order terms will require a prohibitive number of integrations. 

\begin{figure}
\begin{center}
\includegraphics[width=84mm]{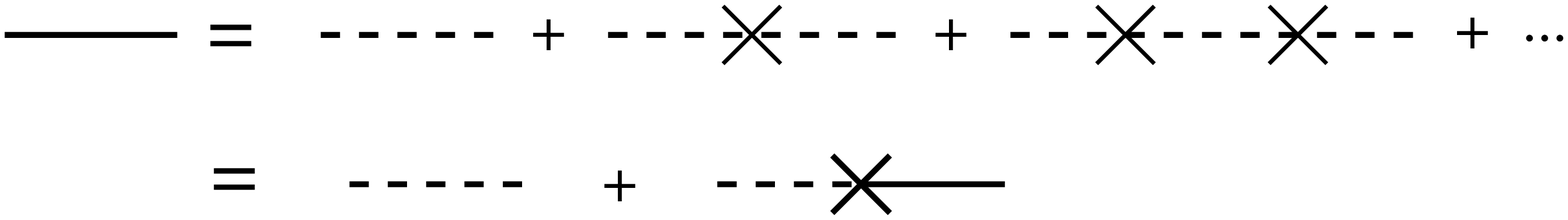}
\end{center}
\caption{Feynman diagrams illustrating equation \ref{equ:summ} for the total probability for a random walk starting from $(S_f,\omega_f)$ to have its first upcrossing between $S_i$ and $S_i+dS_i$ at $\omega_i$. A solid line denotes a period of time in which any sequence of accretion and merger events can take place. A dashed line denotes a period of time in which only accretion takes place and a cross indicates a merger. The first equality says that the total probability is equal to the sum of the fractions of walks which underwent no mergers, one merger, two mergers, etc. We can rearrange the sum to give the second equality, which states that the total probability is equal to the sum of the probability to have no mergers and the probability to have at least one merger.}
\label{fig:feynman}
\end{figure}

Note that the conditional probability densities in Eq.~(\ref{equ:summ}) are analogous to propagators.  It is instructive to visualize each term in equation~\ref{equ:summ} using a diagrammatic notation, as in Fig.~\ref{fig:feynman}.
The figure makes it clear that this equation expands the full propagator in terms of bare propagators (the accretion probability) with interactions (mergers). We can use this insight to rearrange the above equation as in the figure, which shows   that one can resum the terms in Eq.~(\ref{equ:summ}) to write down a tractable integral equation.
 
 Writing the second equality of figure \ref{fig:feynman} in the language of equation \ref{equ:summ}, we find the important result:
\bea
&   &\prob{tot}{i}{i}{f}{f}=\prob{acc}{i}{i}{f}{f} \nonumber \\ +
&   &\int_{\omega_f}^{\omega_i} \!\,d\omega \int_{S_f}^{S_i-\Delta' (S_i)} \! \,dS_1  \int_{S_1+\Delta (S_1)}^{S_i} \,dS_2 \nonumber \\
&   & \prob{acc}{1}{}{f}{f} \probmerge{1}{2}{} \prob{tot}{i}{i}{2}{} \label{equ:main}
\eea
 This is a Voltera integral equation which we will solve numerically. 
 We refer the interested reader for a discussion of solving this equation numerically to appendix \ref{sec:num}. In the next section we compare these semi-analytic results to Monte-Carlo simulations. 
 
\section{comparison with Monte-Carlo simulation}  \label{sec:compare}
   There are several methods for generating a merger tree using Monte-Carlo techniques (see \citet{SK} and \citet{Cole}). Here we choose to use a binary merger tree with accretion method mainly due to its simplicity of implementation. In this scheme the time interval is chosen to be so small that the probability of a merger is very low. This in turn ensures that the probability of more than one merger is negligible in a given time step. Then a halo one time step back will have a smaller mass due to accretion or division into two progenitors. The details of the method can be found in Appendix \ref{sec:monte}.
  
  It should be noted that our Monte-Carlo (MC) simulation gives number weighted probabilities. However, $f_{acc}$ in formula \ref{equ:main} is a mass weighted probability. We can easily change $f_{acc}$ to a number weighted probability $p_{acc}$ using $p_{acc}=\frac{M_{f}}{M} f_{acc}$.

\begin{figure}
\begin{center}
\includegraphics[width=84mm]{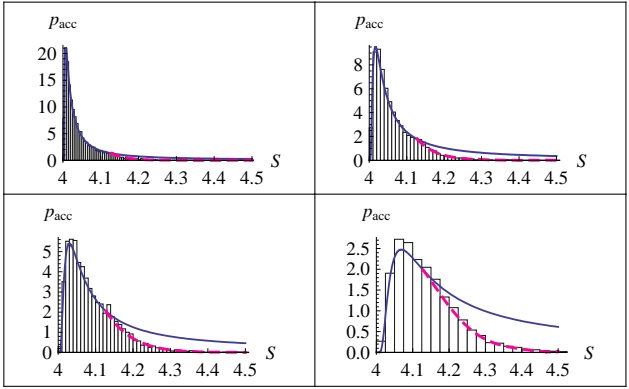}
\end{center}
\caption{This figure shows $p_{acc}$ versus $S_i$, with parent mass $M_f=2 \times 10^{12} M_{\sun}$, mass resolution $M_{res}=\frac{M_{f}}{10}$, and redshift $z_f=0$. The panels show $p_{tot}$ (solid line), our analytic result for $p_{acc}$ (dashed line), and the MC result for $p_{acc}$ (histograms) for increasing redshift from top-left to bottom-right: $z=0.184$, $z=0.267$, $z=0.348$ and $z=0.5$.}
\label{fig:dz}
\end{figure}
  
 Each panel of figure \ref{fig:dz} shows $p_{acc}$ versus $S_i$ for a different lookback redshift $z_i$, with parent mass $M_f=2 \times 10^{12} M_{\sun}$ and redshift $z_f=0$. The mass resolution is fixed at $M_{res}=\frac{M_f}{10}$. The dashed line is our analytical solution, the histogram shows the result of the MC simulation and the solid line is $f_{tot}$ given by equation \ref{equ:ftot}. We see an excellent agreement between the MC simulation and our analytic result, and notice some intuitively sensible trends in the figures that are worth mentioning. First, $p_{acc}=p_{tot}$ for $S_i<4.1$. That is because to have a merger we need  $M_i \le M_f-M_{res}=1.8 \times 10^{12}$, which corresponds to $S_i=4.1$. For $S_i$ smaller than this, the mass jump is always less than mass resolution and only accretion can happen; hence, $p_{acc}=p_{tot}$ in this region. For $S_i$ larger than $4.1$ mergers are allowed, so the probability of having one or more merger is non-zero and $p_{acc}$ will be less than $p_{tot}$. Also, the probability of having at least one merger, $p_{tot}-p_{acc}$, increases monotonically with increasing redshift as more and more halos have a chance to undergo a merger. Finally, as we look further back in time, halos with smaller masses have a chance to reach $M_f$ by just accreting so $p_{acc}$ spreads to smaller masses with increasing redshift.

\begin{figure}
\begin{center}
\includegraphics[width=84mm]{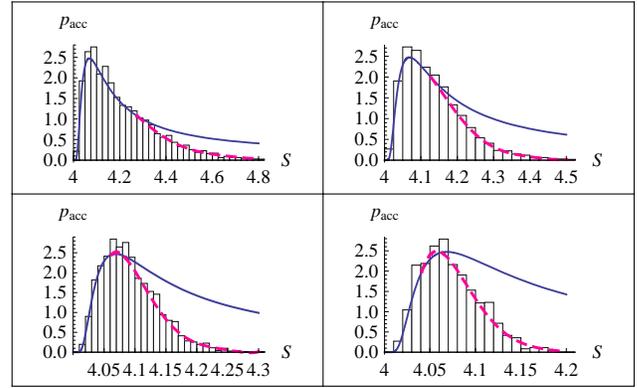}
\end{center}
\caption{The dashed lines in these figures show our analytic result for $p_{acc}$ vs. $S_i$ with the same $M_f$ and $z_f$ as figure \ref{fig:dz}, but here we look at the distribution at a fixed lookback redshift $z_i=0.5$ for decreasing $M_{res}$:  from top-left to bottom-right $M_{res}=\frac{M_{f}}{5}$, $M_{res}=\frac{M_{f}}{10}$, $M_{res}=\frac{M_{f}}{20}$ and $M_{res}=\frac{M_{f}}{30}$. Solid lines show $p_{tot}$ and histograms the results of MC simulations for $p_{acc}$.} 
 \label{fig:dMres}
 
\end{figure}

In figure \ref{fig:dMres} we show $p_{acc}$ vs. $S_i$ with the same $M_f$ and $z_f$ as above, but here we look at the distribution with different choices of $M_{res}$ at a fixed lookback redshift $z_i=0.5$. Again the agreement between our analytical result and the MC simulations is very good. Again, there are some intuitively reasonable trends in the figure that should be noted. As we discussed above, $p_{acc}$ must be equal to $p_{tot}$ for $M_i \ge M_{limit}=M_f-M_{res}$. As $M_{res}$ gets smaller, this limiting mass becomes larger. Therefore $S_{limit}=S(M_{limit})$, the $S_i$ below which $p_{acc}=p_{tot}$, becomes smaller, which can easily be seen in the figure. Notice that different panels have different scales. Also given a fixed lookback redshift, decreasing the mass resolution increases the number of events we classify as mergers thus raising the probability for a halo to have at least one merger. Since $p_{tot}$ is not affected by the choice of $M_{res}$, $p_{acc}$ accordingly decreases with decreasing $M_{res}$. 

\newpage
\section{Scaled solution}  \label{sec:scaled}
In general, for any given prescription for $M_{res}$ one can find the solution of integral equation \ref{equ:main} to obtain $\prob{acc}{i}{i}{f}{f}$. Generally, this needs to be solved for each given final halo mass $M_f$. However, if we impose a special mass resolution for a chosen cosmology it is possible to cast the integral equation into a scale invariant form. Then we need only solve this equation once.
 To achieve this, we need to define $M_{res}$ so as to satisfy the following two equations simultaneously:
 
 \be
 \Delta (S) \equiv \sigma^2 (M(S)-M_{res})-S = C \times S 
 \ee       
 
 and 
 
 \be
 \Delta '(S) \equiv S - \sigma^2(M(S)+M_{res}) = C' \times S
 \ee
where $C$ and $C'$ are constants independent of $S$. Recall that $\Delta(S)$ and $\Delta '(S)$ appear in the limits of integration of equation \ref{equ:main}. For $M_{res}$ small compared to $M(S)$ we can easily see, by Taylor expansion, that to second order in $M_{res}/M$ these equations can be satisfied if we take $C=C'$ and the mass resolution as
 \be
M_{res}(S)=-\frac{S}{dS/dM} C.
 \ee
 For example for a scale invariant matter power spectrum with power index $n$, $P(k) \propto k^n$, the above equation gives
 \be
 M_{res}=\frac{n+3}{3} C M_{parent}
 \label{equ:9}
 \ee
 i.e. a merger is defined when the mass of any progenitor of the halo is larger than a constant fraction of the parent mass\footnote{Note that more general solutions exist if $C\neq C'$ which can possibly be used to loosen the relationship between $M_{res}$ and $M_{parent}$. It seems physically reasonable to choose $M_{res}$ much smaller than and proportional to $M_{parent}$ which is the case when $C=C'$. }.

With this criterion for $M_{res}$ we are ready to rewrite the integral equation \ref{equ:main} in a scale invariant form. To do so we define the new variables

\bea
&  & u \equiv S/S_f \nonumber \\
&  & \theta \equiv (\omega - \omega_f)/S_f^{1/2}
\eea 

and the functions $\tilde{f}_{acc}$ and $\bar{f}_{acc}$:
\bea
 &  &\prob{acc}{1}{}{f}{f} = S_f^{-1} \tilde{f}_{acc}\left(\frac{S_1}{S_f} | \frac{\omega - \omega_f}{S_f^{1/2}}\right) \nonumber \\
 &  & \tilde{f}_{acc} (u_1 | \theta) =\frac{\theta}{(2 \pi)^{1/2} (u_1-1)^{3/2}} \exp{\left(-\frac{\theta^2}{2(u_1-1)}\right)} \bar{f}_{acc}(u_1 | \theta) \nonumber \\
 &  &
 \label{equ:convert}
\eea

With these definitions equation \ref{equ:main} can be written in the manifestly scale invariant form
\bea
1=& &\bar{f}_{acc} (u_i|\theta_i) + \int_{0}^{\theta_i} \!\,d\theta \int_{1}^{u_i(1-C)} \! \,du_1  \int_{u_1(1+C)}^{u_i} \,du_2  \nonumber \\
&  &\bar{f}_{acc} (u_1|\theta)K(\theta_i,u_i,u_1,u_2,\theta)
\label{equ:scaled}
\eea

where the kernel for the spherical collapse model is

\bea
&  & K=\frac{(2 \pi)^{1/2} (u_i-1)^{3/2}}{\theta_i} \exp{\frac{\theta_i^2}{2(u_i-1)}} \times \nonumber \\
&  & \frac{\theta}{(2 \pi)^{1/2}(u_1-1)^{3/2}} \exp{\frac{-\theta^2}{2(u_1-1)}} \times \nonumber \\
&  & \frac{1}{(2 \pi)^{1/2}(u_2-u_1)^{3/2}} \times \frac{(\theta_i -\theta)}{(2 \pi)^{1/2}(u_i-u_2)^{3/2}} \exp{\frac{-(\theta_1-\theta)^2}{2(u_i-u_1)}} \nonumber \\
&  &
\eea
For a given $C$ this equation can be solved for $\bar{f}_{acc}(u|\theta)$. This calculation can be facilitated by noticing that the integral over $S_2$ can be done analytically. Having found $\bar{f}_{acc}(u|\theta)$ one can find $\prob{acc}{1}{1}{f}{f}$ for arbitrary $S_1$, $\omega_1$, $S_f$ and $\omega_f$ using equations \ref{equ:convert}.

\begin{figure}
\begin{center}
\includegraphics[width=84mm]{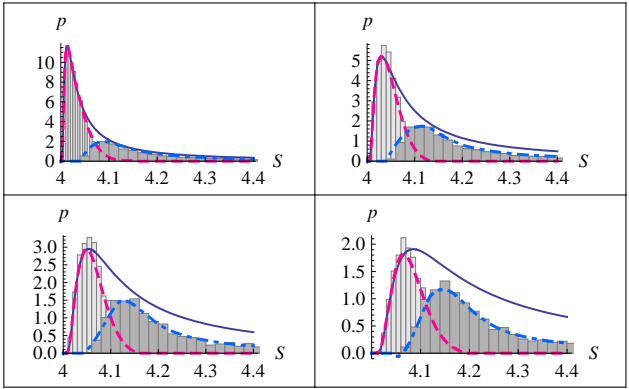}
\end{center}
\caption{Halo accretion and merger probabilities for a power law matter power spectrum with $n=-1$ are shown here. We take $C=0.01$ where $C$ is defined in equation \ref{equ:9}. The solid line denotes $p_{tot}$, the lighter histogram in each panel indicates the result of MC simulation for $p_{acc}$ and the dashed line is our numerical solution of equation \ref{equ:scaled} scaled according to equation \ref{equ:convert}. The darker histogram shows the result of our MC simulation for $p_{1merger}(S_i,\omega_i,S_f,\omega_f)$, the number density of halos in a given range $(S_i,S_i+d S_i)$ that have a parent halo of mass corresponding to $S_f$ at time $\omega_f$ and have merged once in their journey from $\omega_i$ to $\omega_f$. The panels from top-left to bottom right are for $\omega -\omega_f =0.2$, $0.3$, $0.4$ and $0.5$. Assuming $\omega=1.69 (1+z)$ and $z_f=0$ these correspond to lookback redshifts $0.118$, $0.178$, $0.237$ and $0.296$ respectively.}
\label{fig:scaled}
\end{figure}

 The result of this calculation for a power law matter power spectrum with $n=-1$ and $C=0.01$ is shown in figure ~\ref{fig:scaled}. The solid line, as usual, denotes $p_{tot}$, the lighter histogram in each panel indicates the result of MC simulation for $p_{acc}$ and the dashed line on top is our numerical solution of equation \ref{equ:scaled} scaled according to equation \ref{equ:convert}. The panels from top-left to bottom right are for $\omega -\omega_f =0.2$, $0.3$, $0.4$ and $0.5$. Assuming $\omega=1.69 (1+z)$ and $z_f=0$ these correspond to lookback redshifts $0.118$, $0.178$, $0.237$ and $0.296$ respectively. One can see that the agreement is very good for all redshifts considered.
 
 The darker histogram in figure \ref{fig:scaled} shows the result of our MC simulation for $p_{1merger}(S_i,\omega_i,S_f,\omega_f)$. This is the number density of halos in a given range $(S_i,S_i+d S_i)$ that have a parent halo of mass corresponding to $S_f$ at time $\omega_f$ and have merged once in their journey from $\omega_i$ to $\omega_f$. Now that we have found $f_{acc}$ it is possible to calculate $p_{1merger}(S_i,\omega_i,S_f,\omega_f)$ using equation \ref{equ:1merge}. $p_{1merger}$ is nothing but $f_{1merger}$ multiplied by $\frac{M_f}{M}$ to convert from mass density to number density. This result is shown by the dot-dashed line on top of the histogram. The match is very good.
 As expected, for small $\Delta z$ the probability of one merger is much smaller than probability of accretion, which can be seen in the top-left plot. Also in this plot, we can see that the tail of the distribution $p_{1merger}$ approaches $p_{tot}$. This says that the probability of having more than one merger is negligible for small $\Delta z$, as expected.
  
 On the other hand, when $\Delta z$ gets larger more and more halos have a chance to merge, so $p_{acc}$ flattens and $p_{1merger}$ rises. Also, with a large $\Delta z$ there is a finite probability of having more than one merger since the tail of $p_{1merger}$ is considerably below $p_{tot}$. One can continue this calculation and find $p_{2mergers}$, $p_{3mergers}$ and so on, which we have not shown here. Where this hierarchy should be terminated clearly depends on how far we look back in time: a larger $\Delta z$ means more chance of a merger and therefore requires higher merger terms.   
 
\section{discussion and conclusion}  \label{sec:discussion}
 We have used the random walk formalism to find the accretion probability, i.e. the probability for a parent halo to have a progenitor in a given mass interval at a given earlier time given that it has not merged with a halo of a mass larger than the mass resolution. As a concrete example we have worked out the accretion probability in the special case where the barrier is flat, the mass resolution is constant and we look backward in time. However this method can be extended to solve more general problems.  
 
 For example, while we have used a constant barrier, it is well known that this barrier shape does not match the results of N-body simulations. However, our formalism can be generalized to the case of a moving barrier, which has proven to give a very good match to N-body simulations. One only needs to find the appropriate formulae for $f_{tot}$ and $\probmerge{1}{2}{}$ for the moving barrier and solve the integral equation \ref{equ:main}. These functions can in general be found numerically, using for example the method of \citet{Zhang}. However there are analytical results for simple barriers that reproduce the results of N-body simulations, e.g. the square root barrier \citep{Rajesh,Giocoli}. Since the aim of this paper is not to compare with numerical simulations, we will leave this calculation for future work.
 
 
 Here we have always looked backward in time. However, in certain cases it might be more convenient to find $f_{acc}$ in the forward sense. In that case it gives the probability of a halo at an early time being accreted by a larger halo in a given mass interval. This problem can be solved with our formalism by using the forward form of equation \ref{equ:main}, with the forward form of $f_{tot}$ and $f_{merge}$ \citep{Lacey}. Given a population of objects of mass $M$ at time $t$, some of these objects will be destroyed in the course of their evolution by merging with other objects. To find the fraction of objects that have survived from the initial time to the observation time \citep[see][for the case of clusters of galaxies]{Verde} we need to find the fraction  of objects whose halos have not merged with halos more massive than a given threshold; in other words they have only accreted from the initial redshift to the redshift of observation. This is precisely what our formalism calculates.
 
 Finally, this method can lead to a major improvement in the speed of merger tree generation in Monte-Carlo simulations. Since there was no formula for the accretion or merger probabilities in a \emph{finite} time interval, past MC codes had to use \emph{infinitesimal} time-steps to be able to use the known formula for merger probabilities (eqn. \ref{equ:pmerge}), making the computation very time consuming. In this paper we have presented methods to calculate the accretion and merger probabilities for any given time interval which can be used to generate trees with larger time-steps and hence in less computational time. 

\section{Acknowledgment}
EA thanks Laura Book for her comments and aid in manuscript preparation and Akbar Jaefari for his helpful discussion. 
We acknowledge Andrew Benson for his reading of the manuscript and useful suggestions.
BDW acknowledges the Friedrich Wilhelm Bessel prize from the Alexander von Humboldt foundation.  This work was partially supported by an Arnold O.~ Beckman award from the University of Illinois.

%
%
\bibliographystyle{apj} 
\bibliography{ref}
     

\appendix

\section{Brief overview of halo model}  \label{sec:halo}
\begin{figure}
\begin{center}
\includegraphics[width=84mm]{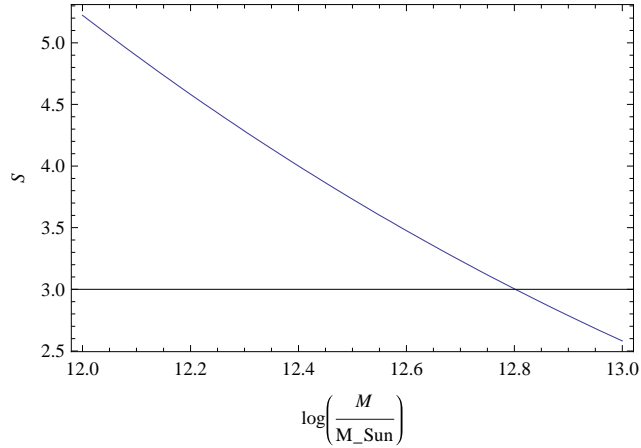}
\end{center}
\caption{This graph shows the rms mass fluctuation inside spheres of mass $M$, i.e $S(M)$, vs mass for our fiducial $\Lambda$CDM cosmology.}
\label{fig:MS}
\end{figure}

In the excursion set formalism one assumes all the matter in the universe is inside collapsed objects, halos, and the aim is to find the number density of these objects for different halo masses. In order to do that one starts from the initial density field of dark matter, which is assumed to be Gaussian, and finds its present time distribution using linear theory. The next step is to take non-linear effects into account, but non-linear theories are generally very complicated and hard to calculate analytically. The halo model circumvents this difficulty by usage of the simplest non-linear model, i.e. the spherical collapse (SC) model, for which a simple analytical solution exists. In the SC model one finds that if an overdense sphere collapses at redshift $z$ then its initial overdensity extrapolated linearly to the present time will have the value $\delta_c D(0)/D(z)$ where $\delta_c$ depends weakly on cosmology and $D$ is the linear growth factor. This fact is used in the model by taking a sphere centered on any point in space and finding these two quantities inside this sphere:
 \be
 \delta=\frac{\rho-\bar{\rho}}{\bar{\rho}}
 \ee
  and  
 \be
 S=\sigma^2 \propto \int k^2 P(k) W^2(k R) dk 
 \ee
 where $\bar{\rho}$ is the mean density of the universe and $\rho$ the mean density inside the sphere. Notice that all of the quantities are calculated using the initial density linearly extrapolated to the present time. $P$ is the matter power spectrum, $W$ is the top-hat window function in real space and $R=(\frac{3M}{4 \pi\bar{\rho}})^{\frac{1}{3}}$. The constant of proportionality is found from $\sigma_8$. For hierarchical cosmologies $S$ is a decreasing function of mass inside the sphere and goes to zero for large radii or equivalently for large masses. Therefore, there is a one to one map between $S$ and $M$. In figure ~\ref{fig:MS}, we give a plot of $S$ as a function of $M$ for the $\Lambda$CDM cosmology with $\Omega_m=0.25$, $h=0.73$ and $\sigma_8=0.9$.
 
  The property that makes this formalism so appealing is that in the case of a Gaussian initial distribution by reducing the radius of the sphere, $\delta$ will execute an uncorrelated random walk for which $S$ is the time like quantity. The plot for $\delta$ versus $S$ for a random point in space will be a realization of a 1-D random walk (see figure \ref{fig:randomwalk}). To find the mass of the halo to which this point belongs at redshift $z$, one draws a horizontal line at $\delta_c(z)$ and finds the point which cuts the random walk for the first time. $S$ at that point shows the mass of the halo to which that point belongs.
  
 With this picture in mind, one can find analytic formulae for halo abundance as a function of their mass: the fraction of volume that belongs to halos in the mass range $(S,S+dS)$ is the fraction of random walks that, starting from $(S=0,\delta=0)$, first cross the barrier of height $\delta_c(z)$ between $(S,S+dS)$. This can be worked out analytically \citep{Bond} and the result is:
\be
 f(S|\delta_c(z)) dS=\frac{1}{\sqrt{2 \pi S}} \frac{\delta_c(z)}{S} \exp \left(-\frac{\delta_c^2(z)}{2 S}\right) dS \label{equ:random1}
\ee

The total mass of the halos in this mass range and in a unit volume will then be $\bar{\rho} f(S|\delta_c(z))$. Finally, the number density of halos is this total mass divided by the mass of an individual halo for the $S$ corresponding to this mass in the $S-M$ relationship (See Figure \ref{fig:MS}). This gives the famous Press-Schechter formula.

 Thinking in terms of random walks has the invaluable advantage of being extendable beyond simple Press-Schechter formulae. \citet{Sheth} used a more realistic model of halos in which they are ellipsoidal objects instead of spherical, and argued that this leads to a barrier that is a function of $S$ for a given redshift. The problem then reduces to a first crossing problem for a moving barrier. The result is in much better agreement with simulations.

\citet{Lacey} argue that changing the origin of the random walk from $(0,0)$ to $(S_f,\delta_c(z_f))$ corresponds to considering only the particles that are inside a halo with mass $S_f$ at redshift $z_f$ and follow their history back in time. Then one can find the mass fraction of halos of mass $S_f$ at redshift $z_f$ that were part of halos within the mass range $(S_i,S_i+dS_i)$ at an earlier redshift $z_i$. This can be achieved by simply changing the origin of the coordinate in equation \ref{equ:random1} from $(0,0)$ to $(S_f,\delta_c(z_f))$ (following \citet{Lacey} we show $\delta_c(z_k)$ as $\omega_k$):
\bea
 &   &\prob{tot}{i}{i}{f}{f}\,dS_i \nonumber \\
 & = & \frac{(\omega_i -\omega_f)}{(2\pi)^{1\!/2}(S_i-S_f)^{3/2}}\exp\left[-\frac{(\omega_i-\omega_f)^2}{2(S_i-S_f)}\right] \,dS_i \label{equ:ptotapp}
\eea 

Putting $\omega_i=\omega_f+d\omega$ and expanding to first order in $d\omega$ gives the probability that the first crossing has a jump from $S_f$ to $S_i$ while one changes the height of the barrier by a tiny $d\omega$ from $\omega_f$ to $\omega_f+d\omega$:
\be
\probmerge{f}{i}{} dS_i d\omega=\frac{1}{(2\pi)^{1\!/2}(S_i-S_f)^{3/2}} dS_i d\omega \label{equ:pmerge}
\ee
Since the probability of multiple mergers is negligible for small $d\omega$, the parent halo can split only into two progenitors during this time interval. 

For more details see \citet{Zentner}

\section{Numerical solution}  \label{sec:num}
The goal of this section is to discretize and solve Eq.~(\ref{equ:main}) for $\prob{acc}{1}{}{f}{f}$ numerically using matrix methods. To start, we rewrite the integral equation for clarity:
\bea
&&\prob{tot}{i}{i}{f}{f}=\prob{acc}{i}{i}{f}{f} + \nonumber \\
&&\int_{\omega_f}^{\omega_i} \!\,d\omega_1 \int_{\omega_f}^{\omega_i} \!\,d\omega_2 \int_{S_f}^{S_i-\Delta ' (S_i)} \! \,dS_1  \int_{S_1+\Delta (S_1)}^{S_i} \,dS_2  \prob{acc}{1}{1}{f}{f} \probmerge{1}{2}{1}\delta(\omega_1-\omega_2) \prob{tot}{i}{i}{2}{2} \nonumber \\
&&    \label{equ:mainn}
\eea

where $f_{acc}$ is unknown, and $f_{merge}$ and $f_{tot}$ are given in equations \ref{equ:pmerge} and \ref{equ:ptotapp} respectively. Also, for a reason which will be clear later we insert an integration of the Dirac delta function into the original equation~\ref{equ:main}. Since we will compare this analytic result with a Monte-Carlo simulation we take $(S_f,\omega_f)$ as given constants which correspond to the mass and redshift of the parent halo.

To solve this equation numerically we discretize $S \in [S_f,S_i]$ into $N_S$ and $\omega \in [\omega_f,\omega_i]$ into $N_{\omega}$ segments. To make sure that discretization is small enough to resolve mergers we demand:
\be
\frac{S_i-S_f}{N_S}\ll \Delta S
\ee
where, like before, $\Delta S$ is defined as the minimum jump in $S$ to have a merge, which for the case we consider here is $\Delta S =S(M(S)-M_{res})-S$.
Also we must make sure that $\Delta \omega$ is small enough for the probability of having more than one merger in that time interval be negligible. For that to be correct we demand:
\be
\frac{\omega_i-\omega_f}{N_{\omega}} \ll \sqrt{\Delta S}
\ee
We will show later that after satisfying these conditions the solution converges by making $\Delta S$ and $\Delta \omega$ smaller and observing that the solution for the integral equation stays the same.

We give a collective index $j$ to any pair $(S_k=k\Delta S,\omega_l=l\Delta \omega)$:

\be
j=(k-1)\times N_S + l \label{eq:above}
\ee 

to go back to original indices $k$ and $l$ we use:

\be
k=[\mod(j-1)]+1	
\ee
and $l$ can be found from this and equation~\ref{eq:above}

Then we define these matrices:
\be
M^{merge}_{i,j} \equiv \left \{ \begin{array}{llcl}
                          \probm{merge}{k_{i}}{l_{i}}{k_{j}}{l_{j}} & \mbox{if $M(S_{k_i})>M(S_{k_j})+M_{res}$ and $\omega_{l_i}=\omega_{l_j}$} \\
                                                   
                          0      \!\!    & \mbox{otherwise}
                          \end{array} 
                          \right.
\ee

\be 
M^{tot}_{i,j} \equiv \left \{ \begin{array}{ll}
                           \prob{tot}{k_i}{l_i}{k_j}{l_j} & \mbox{if $S_{k_i}>S_{k_j}$ and $\omega_{l_i}>\omega_{l_j}$} \\
                           0 & \mbox{otherwise}
                           \end{array}
                           \right. 
\ee
where $i,j \in [1,N_S \times N_{\omega}]$.

Also we define this vector:
\be
V^{tot}_i \equiv \prob{tot}{k_i}{l_i}{f}{f}
\ee
All the above matrices and vectors can be computed numerically. Finally we define the unknown vector:
\be
V^{acc}_i \equiv \prob{acc}{k_i}{l_i}{f}{f}
\ee

Using these definitions we can write the integral equation ~\ref{equ:mainn} in its discretized form as a matrix equation for vector $\boldm{V}^{acc}$:
\be
\boldm{V}^{tot}=\boldm{V}^{acc}+\boldm{M}^{tot} \cdot \boldm{M}^{merge} \cdot \boldm{V}^{acc} \times \left(\Delta S\right)^2 \Delta \omega  \label{equ:matrix1}
\ee  
Notice that the limit of integration is taken into account in the definition of the matrices. Also notice that there is a factor of $\Delta \omega$ instead of $\left(\Delta \omega\right)^2$ in the above formula. That's because there was a delta function in the definition of $f_{merge}$ which gives a factor of $1/\Delta \omega$ in discretization. 
Equation ~\ref{equ:matrix1} has the solution:
\be
\boldm{V}^{acc}=\left(\boldm{1}+\boldm{M}^{tot} \cdot \boldm{M}^{merge} \left(\Delta S\right)^2 \Delta \omega \right)^{-1} \cdot \boldm{V}^{tot}
\ee

Having found $\boldm{V}^{acc}$, we can easily calculate the propagators involving exactly one merger, two mergers, \ldots:
\bea
\boldm{V}^{1merge}&=& \left(\Delta S\right)^2 \Delta \omega \boldm{V}^{acc} \cdot \boldm{M}^{merge} \cdot \boldm{V}^{acc} \nonumber \\
 \boldm{V}^{2merge}&=& \left(\Delta S\right)^2 \Delta \omega \boldm{V}^{1merge} \cdot \boldm{M}^{merge} \cdot \boldm{V}^{acc} \nonumber \nonumber \\
                   &&\vdots
\eea

This completes our numerical solution to the integral equation for the general functions $f_{tot}$ and $f_{acc}$. For the special case of spherical collapse (equations \ref{equ:ptotapp} and \ref{equ:pmerge}) a further simplification is possible by noticing that the first integration over $S_2$ can be done analytically.
 
\section{Monte-Carlo simulation}  \label{sec:monte}

 First we briefly describe how the binary merger with accretion works. For a small change in $\Delta \omega$ the probability of absorbing a mass $\Delta S$ in this time interval is given by:
 \be
 P(\Delta S,\Delta \omega) \,dS =\frac{1}{(2 \pi)^{1/2}}\frac{\Delta \omega}{(\Delta S)^{3/2}}\exp\left[-\frac{(\Delta \omega)^2}{2 \Delta S}\right] \label{equ:extPS}
 \ee
Starting from a parent halo with mass $S_p$ at time $\omega$ we go backward in time to $\omega -\Delta \omega$. Then, if $M(\Delta S)<M_{res}$ we consider that to be accreted mass, stop tracking its history, and take $M_p-M(\Delta S)$ as the new parent halo at redshift $\omega-\Delta \omega$ assuming that this new $M_p$ is larger than $M_{res}$. If not, we consider that as accreted mass and do not continue to track its history.  We choose $\Delta \omega$ small enough to ensure that the probability of having a merger in this time interval becomes small. In other words we demand:
\be
\Delta \omega \ll \sqrt{S(M_p-M_{res})-S(M_p)}
\ee
Then we generate a random number $\Delta S$ consistent with the distribution~\ref{equ:extPS}. This is a very easy task to do since equation~\ref{equ:extPS} can be converted to a Gaussian distribution by a change of variable $x \equiv \Delta \omega/(2\sqrt{\Delta S})$. This procedure is then repeated with the new halos as the parent halos until we reach the time $\omega_i$ where we want to compare our numerical results with the Monte-Carlo simulation. While making the merger tree we keep track of each halo to know how many times in their history they experienced a merger so we will be able to find the distribution of halos with no merger, one merger and so forth.

\end{document}